# Inference based method for realignment of single trial neuronal responses

Tomislav Milekovic[1, 2, 3,*], Carsten Mehring[1, 2, 3]

[1] Bernstein Center Freiburg, University of Freiburg, Hansastr. 9A, 79104 Freiburg, Germany
[2] Faculty of Biology, University of Freiburg, 79104 Freiburg, Germany
[3] Department of Bioengineering and Department of Electrical and Electronic Engineering, Imperial College London, South Kensington Campus, SW7 2AZ London, United Kingdom
[*] current address: Center for Neuroprosthetics and Brain Mind Institute, School of Life Sciences, Swiss Federal Institute of Technology (EPFL), 1015 Lausanne, Switzerland

Corresponding author: Tomislav Milekovic; email: tomislav.milekovic@epfl.ch

**Abstract.** Neuronal responses to sensory stimuli or neuronal responses related to behaviour are often extracted by averaging neuronal activity over large number of experimental trials. Such trial-averaging is carried out to reduce noise and to reduce the influence of other signals unrelated to the corresponding stimulus or behaviour. However, if the recorded neuronal responses are jittered in time with respect to the corresponding stimulus or behaviour, averaging over trials may distort the estimation of the underlying neuronal response. Here, we present an algorithm, named *dTAV* algorithm, for realigning the recorded neuronal activity to an arbitrary internal trigger. Using simulated data, we show that the *dTAV* algorithm can reduce the jitter of neuronal responses for signal to noise ratios of 0.2 or higher, i.e. in cases where the standard deviation of the noise is up to five times larger than the neuronal response amplitude. By removing the jitter and, therefore, enabling more accurate estimation of neuronal responses, the *dTAV* algorithm can improve analysis and interpretation of the responses and improve the accuracy of systems relaying on asynchronous detection of events from neuronal recordings.

# 1. Introduction

Many neurophysiological studies are investigating neuronal responses to external events. These studies range from simple stimulus evoked neuronal responses in the corresponding primary sensory areas, e.g. neuronal responses to light flashes in the primary visual cortex [1], to neuronal activity correlated to complex behaviours, e.g. neuronal correlates of abstract problem solving [2,3]. In such studies, neuronal responses are usually extracted by averaging the neuronal signal in order to reduce the "noise", i.e. parts of the neuronal signal that are not correlated to the stimulus or behaviour that is being investigated. This procedure relies on the assumption that neuronal responses are time locked to the corresponding stimulus or behaviour. This assumption can be challenged, however, as neuronal responses show temporal variability in relation to the corresponding stimulus or behaviour [4-8]. Depending on the amount of the temporal jitter, the underlying neuronal response estimated by averaging may be distorted (Figure 1), possibly leading to mistakes in subsequent analyses and incorrect conclusions.

Here, we propose an inference based algorithm for realignment of neuronal responses ($dTAV$ algorithm). First, we demonstrate that the reduction of variability across trials can be used as a measure of jitter reduction and, therefore, as a measure of how well the neuronal responses have been aligned. This property is crucial for the operation of the $dTAV$ algorithm as it identifies the underlying neuronal response by reducing the variability. Then, we built a simple model of neuronal responses and demonstrated the usability of the $dTAV$ algorithm for various noise levels using simulated data. Furthermore we compare the $dTAV$ algorithm to another non-parametric realignment algorithm [9], referred to as MaxCorr in the rest of the text.

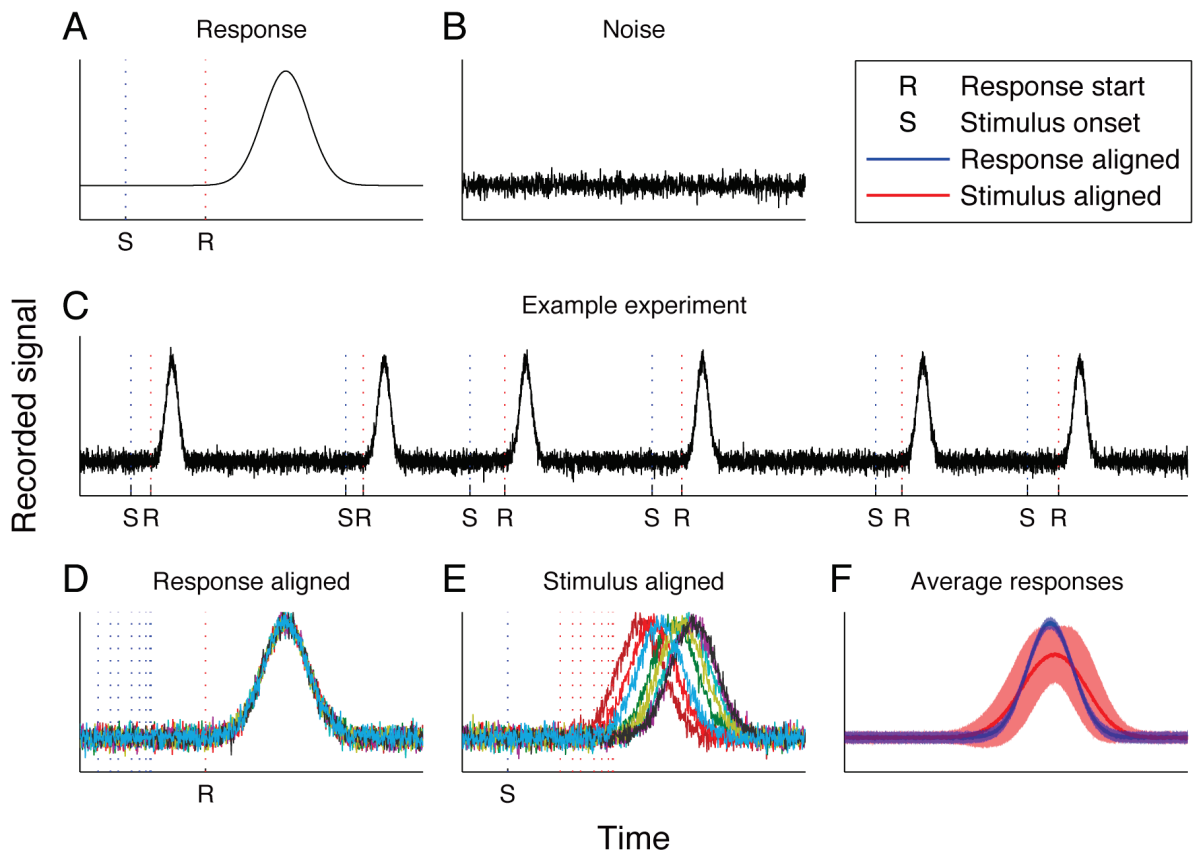

Figure 1. Effect of single trial jitter on the estimation of the underlying neuronal response. A: Neuronal responses are related to the event (E), but are triggered (R) by an internal process, which is not precisely time-locked to the onset of the external event. B: A certain amount of noise is recorded together with the relevant neuronal responses. C: During the experiment, the external event occurs multiple times, while the neuronal activity is recorded. D: When neuronal responses are aligned on the response start, the trial average response (F: blue line) is a good approximation of the real

neuronal response. However, the response onset is unknown. The trial-averaged response aligned on the event onset triggers (F: red line) does not correctly reproduce the real neuronal response. In addition, the standard deviation across trials calculated using the event onset triggers (F: blue and red shaded tubes) is an incorrect estimate of the variability of neuronal responses.

## 2. Methods

The Methods are presented in the following order. First, we describe our simple model of neuronal responses to an external event (stimulus or behaviour). Second, we present the analytic tools used to predict the jitter reduction. Since the exact triggers of neuronal responses are not known in a real-world application of the algorithm, it is necessary to design such a measure of jitter reduction in order to optimize the parameters of the $dTAV$ algorithm. Third, we describe the $dTAV$ realignment algorithm in detail. Finally, we describe the details of simulated data used to assess the performance of the $dTAV$ realignment algorithm and to compare it to a previously published MaxCorr realignment algorithm [9].

### 2.1. A measure of jitter reduction

We assume that the neuronal signal is a superposition of neuronal responses $r(t)$ evoked at response onset times $t_i$ plus the Gaussian white noise signal $\eta$:

$$signal(t) = \sum_i r(t - t_i) + \eta(t) \qquad \eta(t) \in N(0, \sigma_\eta) \tag{1}$$

where $\sigma_\eta$ is the standard deviation of the Gaussian white noise process and $N(\mu, \sigma)$ is a normal distribution with a mean of $\mu$ and a standard deviation of $\sigma$. After recording the neuronal signal and if neuronal response onset times are known, one can estimate the neuronal response by calculating the response-triggered average $\hat{r}(t)$:

$$\hat{r}(t) = \frac{1}{M} \sum_{i=1}^{M} signal(t + t_i) = r(t) + \frac{1}{M} \sum_{i=1}^{M} \eta_i(t) \qquad \eta_i(t) = \eta(t + t_i) \tag{2}$$

$$\hat{s}_\eta(t) = \frac{1}{M} \sum_{i=1}^{M} \eta_i(t) \in N\left(0, \frac{\sigma_\eta}{\sqrt{M}}\right) \tag{3}$$

where $M$ is the number of responses used to calculate the average; $\eta_i(t)$ is the noise in $i$-th trial; and $\hat{s}_\eta(t)$ is the random variable drawn from a Gaussian distribution that follows from the presence of noise. In the following derivations, we will use operator $E(\ )$ for expectation, $V(\ )$ for variance, $\hat{V}(\ )$ for sample variance and $\hat{Cov}(\ )$ for sample covariance. The sample variance of the neural response $r(t)$ in the presence of the noise is given by:

$$\hat{V}(r(t)) = \frac{1}{M-1} \sum_{i=1}^{M} \left(signal(t + t_i) - \hat{r}(t)\right)^2 = \frac{1}{M-1} \sum_{i=1}^{M} \left(\eta_i(t) - \hat{s}_\eta(t)\right)^2 \tag{4}$$

and is distributed as a $\chi 2$ distribution [10]:

$$(M-1) \frac{\hat{V}(r(t))}{\sigma_\eta^2} \sim \chi_{M-1}^2 \tag{5}$$

For a large number of degrees of freedom, i.e. $M$ being large, the $\chi 2$ distribution can be approximated by a normal distribution using the following transformation [11]:

$$a \sim \chi_k^2 \quad \Rightarrow \quad \frac{a - k}{\sqrt{2k}} \sim N(0,1) \tag{6}$$

Using this transformation, equation (5) becomes:

$$\hat{V}(r(t)) = \hat{S}_V(t) \qquad \hat{S}_V(t) \in N\left(\sigma_\eta^2, \sigma_\eta^2 \sqrt{\frac{2}{M-1}}\right) \qquad (7)$$

Experiments are usually designed in such a way that the studied neuronal responses are elicited by events, e.g. sensory stimuli or certain types of behaviour. However, the onset times of the neuronal responses are not known, since these are triggered internally by the brain. The time shift between an event $t_{Ei}$ and the onset of the neuronal response $t_i$ may not be constant and can be regarded as a stochastic process. In our model, the difference between these two time points is modelled by a Gaussian distribution:

$$t_i - t_{Ei} \in N(\mu_J, \sigma_J) \qquad (8)$$

where $\mu_J$ and $\sigma_J$ are the mean and the standard deviation of the distribution. If one has access to event times only, as it is the case in a real experiment where the time points $t_i$ when the brain triggers a response are unknown, one can estimate the neuronal response by calculating the event-triggered sample mean $\hat{r}_J(t)$:

$$\hat{r}_J(t) = \frac{1}{M}\sum_{i=1}^{M} signal(t+t_{Ei}) = \frac{1}{M}\sum_{i=1}^{M} r(t+t_{Ei}-t_i) + \frac{1}{M}\sum_{i=1}^{M}\eta_i(t) = \overline{r}(t) + \hat{s}_\eta(t) \qquad (9)$$

where $\overline{r}(t)$ is the average signal in the presence of the jitter but no noise ($\sigma_\eta=0$). The sample variance of $\hat{r}_J(t)$ is given by:

$$\begin{aligned}
\hat{V}(r_J(t)) &= \frac{1}{M-1}\sum_{i=1}^{M}\left(signal(t+t_{Ei}) - \hat{r}_J(t)\right)^2 \\
&= \frac{1}{M-1}\sum_{i=1}^{M}\left(r(t+t_{Ei}-t_i) - \overline{r}(t) + \eta_i(t) - \hat{s}_\eta(t)\right)^2 \\
&= \frac{1}{M-1}\sum_{i=1}^{M}\left(\eta_i(t) - \hat{s}_\eta(t)\right)^2 + \frac{1}{M-1}\sum_{i=1}^{M}\left(r(t+t_{Ei}-t_i) - \overline{r}(t)\right)^2 \\
&\quad + 2\hat{Cov}\left(r(t+t_{Ei}-t_i), \eta_i(t)\right)
\end{aligned} \qquad (10)$$

The first term of equation (10) is the sample variance $\hat{V}(r(t))$ in the absence of jitter and depends only on the noise $\eta_i(t)$ in a way given by equation (7). The second term is the variance contribution arising from the jitter in the absence of noise and depends only on the jitter. The third term is the sample covariance between the signal and the noise. We can rewrite equation (10) as:

$$\begin{aligned}
\hat{s}_J^2(t) &= \frac{1}{M-1}\sum_{i=1}^{M}\left(r(t+t_{Ei}-t_i) - \overline{r}(t)\right)^2 \\
\hat{S}_{Cov}(t) &= 2\hat{Cov}\left(r(t+t_{Ei}-t_i), \eta_i(t)\right) \\
\hat{V}(r_J(t)) &= \hat{S}_V(t) + \hat{s}_J^2(t) + \hat{S}_{Cov}(t)
\end{aligned} \qquad (11)$$

For normally distributed and small jitters $t_i - t_{Ei}$, we can use a Taylor series expansion to express $\hat{s}_J^2(t)$:

$$\bar{r}(t) = \frac{1}{M}\sum_{i=1}^{M} r(t+t_{Ei}-t_i) \approx \frac{1}{M}\sum_{i=1}^{M}\left(r(t)+(t_{Ei}-t_i)\frac{dr(t)}{dt}\right) = r(t)+\frac{dr(t)}{dt}\frac{1}{M}\sum_{i=1}^{M}(t_{Ei}-t_i)$$

$$\hat{s}_J^2(t) = \frac{1}{M-1}\sum_{i=1}^{M}\left(r(t+t_{Ei}-t_i)-\bar{r}(t)\right)^2$$

$$\approx \frac{1}{M-1}\sum_{i=1}^{M}\left(r(t)+(t_{Ei}-t_i)\frac{dr(t)}{dt}-r(t)-\frac{dr(t)}{dt}\frac{1}{M}\sum_{j=1}^{M}(t_{Ej}-t_j)\right)^2 \quad (12)$$

$$\approx \frac{dr(t)}{dt}\frac{1}{M-1}\sum_{i=1}^{M}\left((t_{Ei}-t_i)-\frac{1}{M}\sum_{i=1}^{M}(t_{Ei}-t_i)\right)^2$$

$$\approx \frac{dr(t)}{dt}\hat{V}(t_{Ei}-t_i) \;\Rightarrow\; \frac{(M-1)\,\hat{s}_J^2(t)}{\frac{dr(t)}{dt}\,\sigma_J^2} \sim \chi^2_{M-1}$$

For large $M$, we can use transformation (6) and obtain:

$$\hat{s}_J^2(t) \in N\left(\frac{dr(t)}{dt}\sigma_J^2, \frac{dr(t)}{dt}\sigma_J^2\sqrt{\frac{2}{M-1}}\right) \quad (13)$$

Since $r(t + t_{Ei} - t_i)$ and the noise $\eta$ are not correlated, the expectation of $\hat{S}_{Cov}(t)$ is zero while its variance depends on the shape of the neuronal response and, thus, cannot be precisely estimated.

Using equations (7) and (12), the expectation of $\hat{V}(r_J(t))$ for large M and small jitters can be expressed as:

$$\mathrm{E}\left(\hat{V}(r_J(t))\right) = \sigma_\eta^2 + \frac{dr(t)}{dt}\sigma_J^2 \quad (14)$$

We can now see from equation (14) how comparing $\hat{V}(r_J(t))$ values may be used to infer which one of the two $\sigma_J$ values, e.g. $\sigma_J'$ and $\sigma_J''$, is larger than the other. If $\sigma_J' > \sigma_J''$, the corresponding expectations of the $\hat{V}(r_J(t))$, $\hat{V}(r_J(t))'$ and $\hat{V}(r_J(t))''$, will follow this relationship:

$$\sigma_J' > \sigma_J'' \;\Rightarrow\; \mathrm{E}\left(\hat{V}(r_J(t))'\right) > \mathrm{E}\left(\hat{V}(r_J(t))''\right) \quad (15)$$

To optimize the alignment, it is necessary to reduce the amount of jitter without the knowledge of the real neuronal response triggers $t_i$. A possible approach minimizes the variance of the stimulus triggered response $\hat{V}(r_J(t))$ (equation 10) at one particular time point since a reduction of jitter may result in a decrease of $\hat{s}_J$ and, hence, lead to a smaller variance (equation 16). However, since the $\hat{S}_V$ and $\hat{S}_{Cov}$ terms depend on the noise in the signal and, thus, come from a stochastic process, their values might go up by chance and, therefore, mask the reduction of $\hat{s}_J$. Neighbouring time points of the neuronal response are correlated in time and so is the variance term $\hat{s}_J$ arising from the jitter. On the other hand, the noise may be correlated on a smaller time scale. Therefore, the stochastic values of $\hat{S}_V$ and $\hat{S}_{Cov}$ may average out across time if the variance is averaged across a sufficiently large time window. A more reliable indicator of jitter reduction may, therefore, be the decrease of the time-averaged variance, *TAV*:

$$TAV = \langle \tilde{V}(r_J(t)) \rangle = \frac{1}{(T_E - T_S)} \int_{T_S}^{T_E} \tilde{V}(r_J(t)) dt \qquad (16)$$

As discussed before, averaging over time can reduce the variance of the $TAV$, thereby increasing the reliability of the measure. On the other hand, integrating over periods of time where the neuronal response is small compared to the noise or completely absent may not increase the reliability of $TAV$ but instead increase the variance of $TAV$. This trade-off means that the integration window across which the variance is averaged should neither be too long nor too short.

In this study, we used the difference of $TAV$, $dTAV$, as a measure of jitter reduction:

$$dTAV(\sigma'_J, \sigma''_J) = TAV(\sigma'_J) - TAV(\sigma''_J) \qquad (17)$$

We used dTAV to optimize parameters of our realignment algorithm by assuming the following is true:

$$dTAV(\sigma'_J, \sigma'''_J) > dTAV(\sigma'_J, \sigma''_J) \quad \Rightarrow \quad p(\sigma'''_J < \sigma''_J) > p(\sigma'''_J > \sigma''_J) \qquad (18)$$

If equation (19) is correct, we can choose those parameters of our realignment algorithm which lead to the strongest decrease of the time-averaged variability, i.e. we select the parameters for which $dTAV$ is largest. While equations (16) and (17) indicate that equation (19) is correct, analytical derivation of such relationship will depend on the neural responses and may not always hold. In the next section, we use numerical simulations to show that such relationship holds for two simulated neural responses that have a mono-phasic and a bi-phasic shape.

## 2.2. Numerical analysis of the reliability of *dTAV* as a measure of jitter reduction

To demonstrate the conditions for which the $dTAV$ is a reliable measure of the reduction in jitter, we performed a set of simulations, each composed of 2000 repetitions of an experiment composed of 100 trials. The neuronal responses were simulated as mono-phasic and bi-phasic functions composed from Gaussian functions (Figure 2):

$$r_{mono}(t) = \frac{1}{\sqrt{2\pi}\sigma_R} e^{-\frac{t^2}{2\sigma_R^2}} \qquad (19)$$

$$r_{bi}(t) = \frac{1}{\sqrt{2\pi}\sigma_R} e^{-\frac{t^2}{2\sigma_R^2}} - 1.5 \cdot \frac{1}{\sqrt{2\pi} \cdot 3\sigma_R} e^{-\frac{(t-5\sigma_R)^2}{2 \cdot (3\sigma_R)^2}} \qquad (20)$$

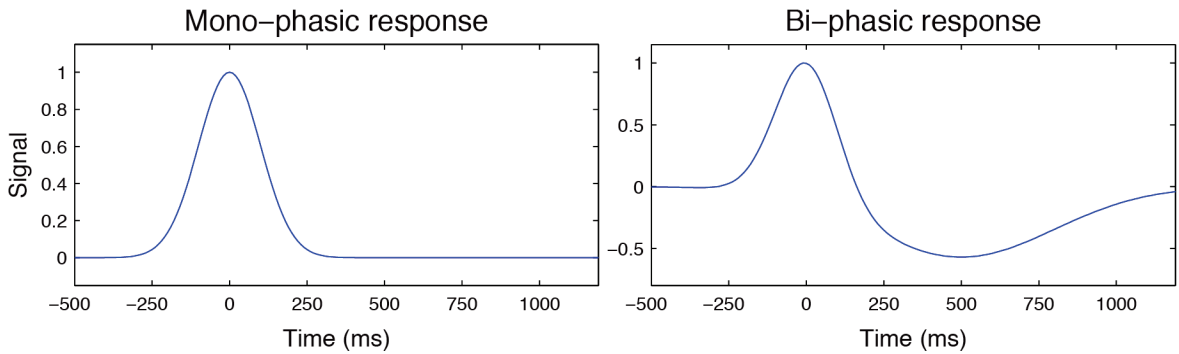

**Figure 2. Simulated mono-phasic (left) and bi-phasic (right) neuronal responses.**

where $\sigma_R$ was taken to be 100ms. The shifts of the neural responses, $t_i - t_{Ei}$, were drawn from a Gaussian distribution with zero mean and standard deviation $\sigma_J$:

$$t_i - t_{Ei} \in N(0, \sigma_J) \qquad (21)$$

Noise was modelled as white Gaussian noise with zero mean and standard deviation $\sigma_\eta$ (equation 1). Our simulations used discrete time with a time step of 1ms. The $TAV$ was calculated for each combination of $\sigma_J$, $\sigma_\eta$ and the integration time $T_I$; and for each simulation run $k$ using the following equation:

$$TAV_k(T_I, \sigma_\eta, \sigma_J) = \frac{1}{2T_I + 1} \sum_{t=-T_I}^{T_I} \tilde{V}_k(r_J(t; \sigma_\eta, \sigma_J)) \qquad (22)$$

Each simulation was performed by selecting a combination of $\sigma_J$, $\sigma_\eta$ and $T_I$ values. The used $\sigma_J$ values ranged from 0ms to 120ms in steps of 1ms and the simulation for $\sigma_J = 60$ms was performed twice because a dataset with 60ms jitter was used as the starting point for the simulated realignment. $T_I$ values ranged from 30ms to 990ms in steps of 30ms. The $\sigma_\eta$ values were selected to model different signal to noise ratios (SNRs), defined as the ratio of the maximum absolute value of the neuronal response and the standard deviation of the noise $\sigma_\eta$. We used $\sigma_\eta$ values that yielded SNR values of 0.03, 0.05, 0.08, 0.13, 0.20, 0.32, 0.50, 0.79, 1.26 and 2.00 for both mono and bi-phasic responses.

We used these simulations to emulate an experiment where the jitter standard deviation of the dataset was $\sigma_J' = 60$ms before the realignment. This initial dataset was compared to datasets with jitter standard deviations $\sigma_J''$ ranging from 0ms (no jitter) to 120ms (doubled jitter) which represented the dataset after the realignment. $dTAV$ was then calculated for each combination of $k$, $\sigma_J''$, $\sigma_\eta$ and $T_I$.

Ranges of $dTAV$ values varied across different orders of magnitude for different $\sigma_\eta$ and $T_I$ values. We therefore normalized $dTAV$ values by dividing them by the maximum of the absolute value of the 1st and 99th percentile.

$$\begin{aligned}
dTAV_{1\%}(T_I, \sigma_\eta) &= \underset{k, \sigma_J''}{\overset{1\%}{P}}\left(dTAV_k(T_I, \sigma_\eta, \sigma_J', \sigma_J'')\right) \\
dTAV_{99\%}(T_I, \sigma_\eta) &= \underset{k, \sigma_J''}{\overset{99\%}{P}}\left(dTAV_k(T_I, \sigma_\eta, \sigma_J', \sigma_J'')\right) \\
dTAV_N(T_I, \sigma_\eta) &= \max\left(|dTAV_{1\%}(T_I, \sigma_\eta)|, |dTAV_{99\%}(T_I, \sigma_\eta)|\right) \\
ndTAV(T_I, \sigma_\eta, \sigma_J', \sigma_J'') &= \frac{dTAV_k(T_I, \sigma_\eta, \sigma_J', \sigma_J'')}{dTAV_N(T_I, \sigma_\eta)}
\end{aligned} \qquad (22)$$

where $\underset{a,b}{\overset{X\%}{P}}$ is the $X$-th percentile operator acting over variables $a$ and $b$; and max is the maximum value operator. The normalized $dTAV$ ($ndTAV$) was binned in 50 equally wide bins spanning the space from -1 to 1. Binned values were used to calculate the probability of jitter reduction, $p(\sigma_J' > \sigma_J'')$, for different $ndTAV$ values, while keeping $\sigma_\eta$ and $T_I$ constant. To show how the reliability of $dTAV$ as a measure of jitter reduction depends on SNR and $T_I$, we calculated the $p(\sigma_J' > \sigma_J'' | ndTAV > 0) = 0.9$ contours in the space spanned by $ndTAV$ and $T_I$ for each value of SNR separately. We also calculated the joint probabilities for each combination of $\sigma_\eta$ and $T_I$ values, $p((\sigma_J' - \sigma_J'')/\sigma_J', ndTAV)$, in order to verify that the relationship in equation (19) holds.

## 2.3. *dTAV* realignment algorithm

The *dTAV* realignment algorithm (Figure 3) relies on the assumption that the distribution of shifts in the recorded neuronal signal is unimodal and that the neuronal responses can be represented by a small number of features ($f_1,...,f_n$). In our case, these features were neuronal signals recorded at different equidistant time points around the time of the event ($\tau_1,...,\tau_n$):

$$f_{k,i} = signal(t_{Ei} + \tau_k) \tag{23}$$

In the first step, we parameterized the ($\tau_1,...,\tau_n$) set by the following three parameters: (i) time of the first feature $t_1$, (ii) number of features $n$ and (iii) temporal distance between the last and the first feature $t_n - t_1$.

The second step of our *dTAV* realignment algorithm is to select a subset of trials $S$ that are already fairly well aligned (Figure 3b). We selected a subgroup of trials containing half of the total number of trials that has a small variance in the Euclidean space spanned by the features when compared to other such subsets. The selection was performed using an iterative selection algorithm that with initited with one trial, $\alpha_1$, and then consecutively adds a trial to the set of selected trials until this set contains half of the trials. The selection process operates as follows: Let Ψ be the set of all trials and Θ be the set of selected trials. In one iteration the trial that is closest to the mean of the set of selected trials is added to the set:

$$F_l = \begin{bmatrix} f_{1,l} \\ \vdots \\ f_{n,l} \end{bmatrix}, \quad \mu_k = \frac{1}{k}\sum_{i=1}^{k} F_{\alpha_i} = \begin{bmatrix} \frac{1}{k}\sum_{i=1}^{k} f_{1,\alpha_i} \\ \vdots \\ \frac{1}{k}\sum_{i=1}^{k} f_{n,\alpha_i} \end{bmatrix} \tag{24}$$

$$\alpha_{k+1} = \min_l \arg \|F_l - \mu_k\|_2, \quad F_l \in \Psi / \Theta \tag{25}$$

This iterations proceeds until half of the trials have been selected. We then calculated the variance $V(\alpha_1)$ of this set. The algorithm is then started again, this time using a different initial trial. This procedure is repeated until all trials have been used once to initiate the selection algorithm. The subset $S$ with the smallest variance was selected:

$$S = \min_i \arg(V(\alpha_i)) \quad \alpha_i \in \Psi \tag{26}$$

In the third step (Figure 3c), the selected trials were used to build a sliding window detection model. Features of the selected trials constituted the "response" class ($R_{class}$). The "baseline" class ($B_{class}$) was constituted from features taken from the same trial subset at times that differed from ($\tau_1,...,\tau_n$) by integer multiples of 1ms.

$$\begin{aligned} R_{class} &= \{(f_{1,i},...,f_{n,i}), i \in S\} \\ B_{class} &= \{(signal(t_{Ei} + \tau_1 + \Delta t),...,signal(t_{Ei} + \tau_n + \Delta t)), i \in S, \Delta t = k \cdot 1ms, k \in \mathbb{Z}/\{0\}\} \end{aligned} \tag{27}$$

These two classes were then used to train a quadratic discriminant analysis (QDA) [12] by fitting Gaussian distributions to each of the classes. The QDA was then applied to neural feature vectors to calculate the probabilities that the feature vectors belong to one of the classes.

In the next step (Figure 3d), the QDA model is used to give a probability $p_R$ that a feature set from a given trial belongs to the response class. For each of the trials, we calculated the time of the maximum probability in a certain time range:

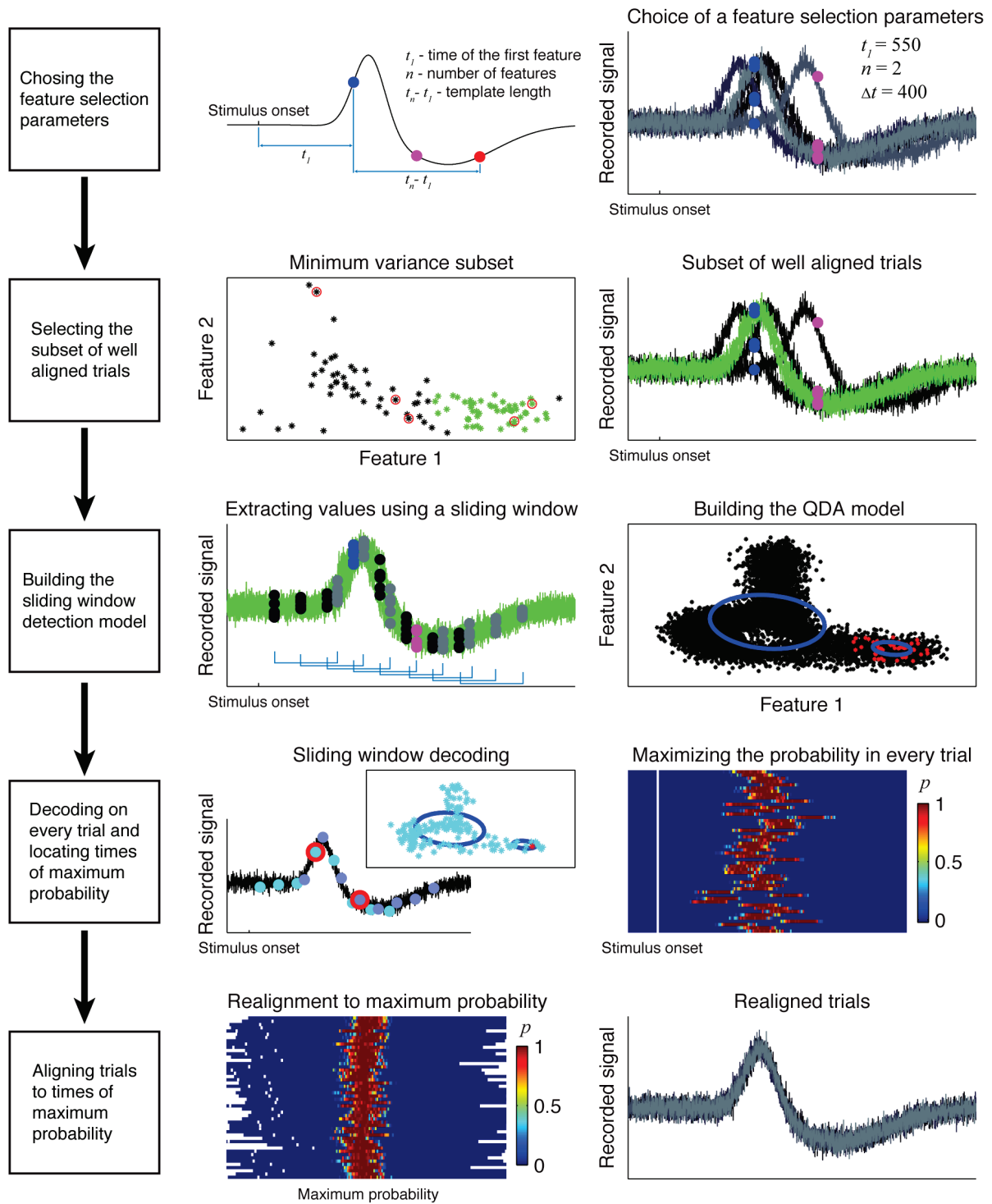

**Figure 3. Processing steps of the *dTAV* realignment algorithm. First step (A):** The response is represented by a small number of features. Left: The selected features are determined by the chosen array of equidistant time points parameterized by the time of the first feature $t_1$, the number of features $n$ and the temporal distance between the last and the first feature $t_n - t_1$. Right: Five example trials and the values of the chosen response features (blue and magenta circles). **Second step (B):** The subset of half of the trials with the smallest within-subset variance is selected. Left: The values of the chosen response features for every trial shown in the feature space. Green stars show the chosen subset. Right: Five example trials with trials belonging to the chosen subset shown in green. Red circles in the left panel show the feature values of the five example trials. **Third step (C):** The QDA model is built from the population of values extracted by the time sliding template. Left: Feature values used to calibrate the model are taken only from the selected subset of trials. Right: The response class is made by extracting features using the previously chosen array (left: blue and magenta dots; right: red stars), while the "baseline" class is made by sliding the same array in time (left: first feature - black dots; second feature - grey dots at the temporal distance of $t_n$- $t_1$ from the black dots; light blue brackets connect first and second features of a pair; right: feature pairs represented by black stars). Blue ellipses show the estimated standard deviation for the "response" and "baseline" class. Fourth step (D): The posterior probability of belonging to the "response" class was calculated by sliding the chosen array in every trial. Left: Extracted features in one trial and in feature space. Red circle and red star represent the feature with the maximum probability for the "response" class. Right: Features with maximum probability for "response" class were found in each trial. Fifth step (E): Left: Trials were realigned to the points of maximum probability. Right: Five example trials realigned to the point of maximum probability.

$$t_{i,MAX} = \max_{t} \arg \left( p_R \left( \begin{bmatrix} signal(t_{Ei} + \tau_1 + t) \\ \vdots \\ signal(t_{Ei} + \tau_n + t) \end{bmatrix} \in R_{class} \right), t_S < t < t_E \right) \quad (28)$$

and used it to re-align the stimulus-triggered neuronal responses and calculate $dTAV$ (Fig. 2e):

$$\breve{r}(t) = \frac{1}{M} \sum_{i=1}^{M} signal(t - t_{Ei} - t_{i,MAX}) \quad (29)$$

$$dTAV(\tau_1,...,\tau_n) = TAV_{START} - TAV(t_{1,MAX},...,t_{M,MAX}) \quad (30)$$

This procedure was repeated for a different selection of features representing the neuronal response.

In addition to parameters that determine feature extraction, it is possible to modify other parameters in the realignment algorithm, such as the integration time used to calculate the $TAV$ or the size of the trial subgroup used to build the QDA model.

To train the QDA, a larger or smaller trial subset could be used. However, if the number of selected trials is small, it will be difficult to reliably estimate the Gaussian distributions. On the other hand, if the number of selected trials is very large and close to the total number of trials, the QDA will be trained on many not well aligned trials and therefore, not be able to reduce the jitter. We assumed that using half of the trials may be a good compromise but this value may be adjusted when the method is applied to other datasets.

After all the parameter values have been exhausted, a set of time shifts was chosen by maximizing $dTAV$:

$$(t_{1,MAX},...,t_{M,MAX})_{Chosen} = \max_{parameters} \arg(dTAV) \quad (31)$$

### 2.4. Simulated data

We used $dTAV$ realignment algorithm to realign the neuronal responses in a range of simulated experiments. In addition to $dTAV$ algorithm, we also used the MaxCorr algorithm [9] and compared the results of the two algorithms.

Simulations were made for two neuronal responses, mono-phasic ($r_M$) and bi-phasic responses ($r_B$), whose shape resembled reported neurophysiological responses. For both types of responses, we performed 100 simulated experiments, each consisting of 200 trials. In each trial, the single channel neuronal response to an arbitrary stimulus was recorded at 1KHz. Neuronal responses were modelled as follows (Figure 2):

$$r_M(t) = \begin{cases} e^{\frac{(t-250)^2}{2 \cdot (83)^2}} & \text{for } 0 \leq t < 500 \\ 0 & \text{otherwise} \end{cases} \quad (32)$$

$$r_B(t) = \begin{cases} e^{\frac{(t-125)^2}{2 \cdot (25)^2}} - 1.5 e^{\frac{(t-250)^2}{2 \cdot (83)^2}} & \text{for } 0 \leq t < 500 \\ 0 & \text{otherwise} \end{cases} \quad (33)$$

Neural responses were simulated in the following way:

$$data_i(t) = \sum_j r\left(t - t_j + \left(t_j - t_{Ej}\right)\right) + \eta(t) \quad (34)$$

$$\eta(t) \in N(0, \sigma_{\eta,i}); \; t_{j+1} - t_j \in N(10s, 10s); \; t_j - t_{Ej} \in N(0, 0.1s)$$

Noise in the recordings was simulated as additive Gaussian noise with zero mean and different standard deviations $\sigma_\eta$: 31.62, 19.95, 12.59, 7.94, 5.01, 3.16, 2.00, 1.26, 0.79 and 0.50, which corresponded to SNRs of 0.03, 0.05, 0.08, 0.13, 0.20, 0.32, 0.50, 0.79, 1.26 and 2.00. Temporal distances between the stimulus times were drawn from a Gaussian distribution with a mean of 10s and a standard deviation of 10s. To keep neuronal responses from overlapping, temporal distances below 3s were redrawn. The neuronal response offset (i.e. the temporal jitters) were drawn from a Gaussian distribution with zero mean and a standard deviation of 0.1s. To avoid occasional very large jitters, all jitters with an absolute value above 0.3s were redrawn.

To correctly simulate the outcome of the experiment, we assumed that the person analysing the data would filter the data using a low-pass filter, given that low-frequencies dominate the simulated neuronal responses. We filtered the simulated recordings using 2$^{nd}$ order symmetric Savitzky-Golay filters [13,14] with different time windows of 100ms, 250ms, 500ms or 1000ms.

The time points used to extract the neuronal features were varied using three parameters: time of the first feature relative to the time of the event $\tau_1$ (values ranged from -125ms to 1324ms in steps of 63ms), temporal distance between the first and the last feature $\tau_1$- $\tau_n$ (100ms, 250ms, 500ms or 1000ms) and the number of features $n$ (2, 4, 8 or 12). We used an integration time $T_I$ of 350ms and the maximum probability was identified in the time window ranging from 300ms before the stimulus till 300ms after the stimulus.

### 2.5. MaxCorr realignment algorithm

Realignment results obtained using $dTAV$ algorithm were compared to results obtained using the MaxCorr algorithm [9]. The MaxCorr algorithm works by approximately maximizing cross-correlations between each pair of trials in three steps. First, for $N$ trials, $N(N-1)/2$ cross-correlations $CX_{ij}$ for all possible trial pairs $i$ and $j$ and time lags $(\lambda_i - \lambda_j)$ up to half of the trial length are calculated. Second, a parabolic function $F_{ij}(\lambda_i - \lambda_j)$ is fitted to the crosscorrelation between the $i$-th and the $j$-th trial around the time lag $(\lambda_i - \lambda_j)_{MAX}$ for which the crosscorrelation is at maximum.

$$CX_{ij}\left(\lambda_i - \lambda_j\right) \approx F_{ij}\left(\lambda_i - \lambda_j\right) = b_0 + b_1\left(\lambda_i - \lambda_j\right) + b_2\left(\lambda_i - \lambda_j\right)^2 \quad (35)$$

To fit the parabolic functions $F_{ij}(\lambda_i - \lambda_j)$, we used a neighbourhood of ±10ms around $(\lambda_i - \lambda_j)_{MAX}$. We tested other neighbourhoods of up to 60ms and found that the size of the neighbourhood did not influence the results (data not shown). In the last step, all parabolic functions are summed up to derive a new function $F(\lambda_2, ..., \lambda_N)$, which is quadratic in all of its variables and, therefore, has a unique global maximum $(\lambda_2, ..., \lambda_N)_{MAX}$. The values of $(\lambda_2, ..., \lambda_N)_{MAX}$ are calculated by solving the system of linear equations obtained by applying partial derivatives to the parabolic function $F$. $(\lambda_2, ..., \lambda_N)$ are used to realign the trials relative to the first trial. MaxCorr algorithm was applied to the same signal used to evaluate the realignment of the $dTAV$ algorithm and we evaluated jitter reduction metrics $(\sigma_J' - \sigma_J'') / \sigma_J'$ for both algorithms for comparison.

# 3. Results

## 3.1. *dTAV* as a measure of jitter reduction

Using the model of neuronal responses described in section 2.1, we calculated the dependence of *dTAV* on the reduction of the jitter for different SNR levels and integration times $T_I$. Results are summarized in Figure 4 for the mono-phasic signal and in Figure 5 for the bi-phasic signal.

For an integration time of $T_I$=300ms, selected for presentation in Figures 4a-c and 5a-c, the expectation of the *dTAV* increased monotonously with the amount of reduction of the jitter and did hardly depend on the SNR (Figure 4a, Figure 5a). On the other hand, the standard deviation of *dTAV* (Figure 4b, Figure 5b) depended strongly on the SNR and was comparable or larger than the expected value of the *dTAV*. For high SNRs, the expectation of *dTAV* surpassed the standard deviation of *dTAV* even for small reduction in jitter (see Figure 4a-b and 5a-b). In such cases, *dTAV* is a reliable measure of the amount of jitter reduction, even when the amount of jitter reduction is small.

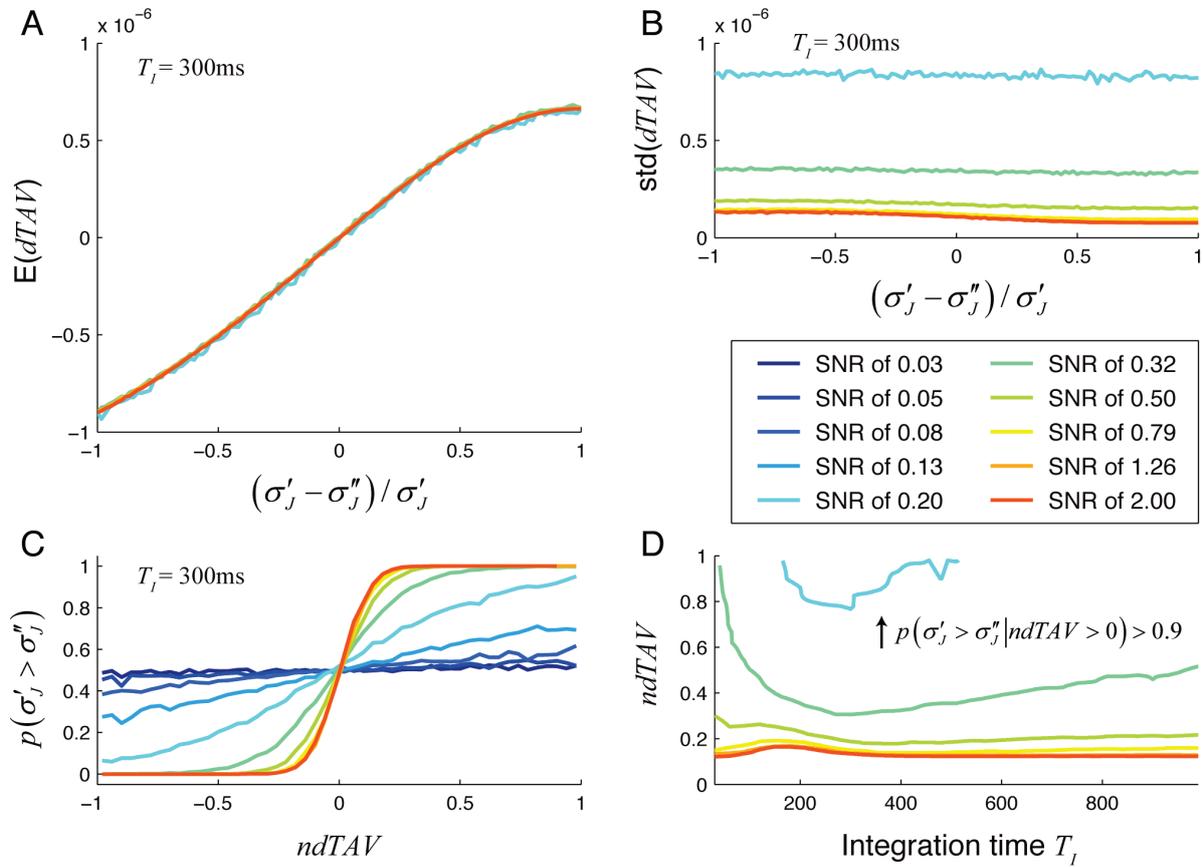

Figure 4. Reliability of *dTAV* as a measure of jitter reduction for mono-phasic neuronal responses. A: Expectation of *dTAV* as a function of the reduction of jitter standard deviation. Lines drawn only for values of SNR of 0.2 and higher. For lower SNR, 2000 repetitions were insufficient to provide a reliable estimate of the expected value of *dTAV* due to the high noise level. For the shown SNR range, the expected value of *dTAV* is independent of the SNR. B: The standard deviation (std) of *dTAV* as a function of the amount of jitter reduction for different SNRs. Standard deviations of *dTAV* for SNR of 0.2 and lower are above $10^{-6}$ and are, therefore, not shown. C: Probability of jitter reduction as a function of *ndTAV* for different SNRs. Panels A, B and C are shown for integration time $T_I$ of 300ms. D: Values of jitter reduction and integration times for which the probability of correct *dTAV* prediction reaches 90%. For jitter reductions and integration times above the line, the probability for correct *dTAV* prediction, $p(\sigma'_J > \sigma''_J | ndTAV > 0)$, is above 90%.

For SNRs of 0.13 and lower, the probability of correct *dTAV* prediction never reached 90%.

To measure how well we can rely on the $dTAV$ as a measure of jitter reduction, we calculated the probability of correctly predicting that the jitter was reduced based on the normalized $dTAV$ values (Figure 4c, Figure 5c). As the SNR was increased, the probability increased up to 1, even for the smallest jitter reductions. For low SNR values the probability never reached 1, even when the jitter was completely removed. To reach a substantial increase of the probability above 0.5, an SNR of about 0.20 or higher was needed.

To provide an insight into the dependence of the probability of correct $dTAV$ prediction on the integration time, we calculated the values of integration times and reductions of jitter standard deviation for which the probability of correct $dTAV$ prediction reached 90% (Figure 4d, Figure 5d). For high SNRs the performance of the $dTAV$ prediction was nearly independent on the integration time whereas for low SNRs the integration time had a stronger influence on the performance of the $dTAV$ prediction. The performance of the $dTAV$ prediction decreased faster for integration times below the optimal integration time, while it decreased more slowly for integration times bigger than the optimal integration time. Therefore, choosing a short integration time could be more disadvantageous than choosing a longer integration time.

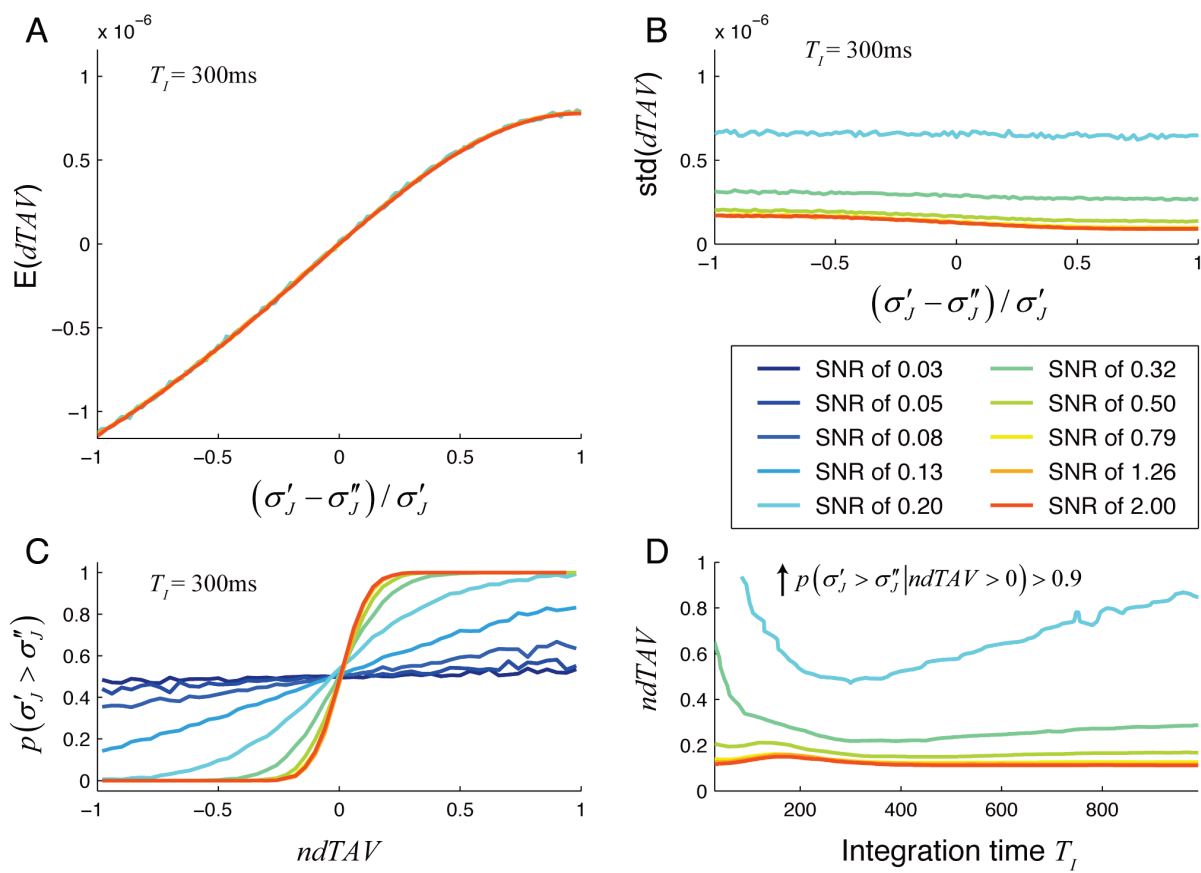

Figure 5. Reliability of $dTAV$ as a measure of jitter reduction for bi-phasic neuronal response. See caption of Figure 4 for details.

So far we have computed the probability of jitter reduction given a reduction in $dTAV$. Next, we investigated whether the difference of $dTAV$ is also predictor of the expected amount of jitter reduction, i.e. does a larger reduction in $dTAV$ also indicate a larger amount of jitter reduction. If so, we can use $dTAV$ to optimize parameters of our re-alignment algorithm by selecting parameter values that gave the highest $dTAV$ values. To assess whether this is the case, we calculated the joint probability distribution of jitter reduction and $dTAV$ for different SNR values for the mono-phasic neural response (Figure 6). For SNRs of 0.13 and lower, $dTAV$ provided no or only very little information of the amount of jitter reduction but as the SNR increased (to values of about 0.2 and higher), the relation between $dTAV$ and jitter reduction became less variable and $dTAV$ became an increasingly good predictor of the amount of jitter reduction. For high SNRs (1.26 and higher), even small differences in $dTAV$ indicated increased jitter reduction with high certainty. This suggests that $dTAV$ is a good predictor of the amount of jitter reduction for sufficiently high SNRs (of about 0.2 and higher) and can be used to optimize the parameters of our realignment algorithm in such cases. The parameter selection based on $dTAV$ will improve with increasing SNR.

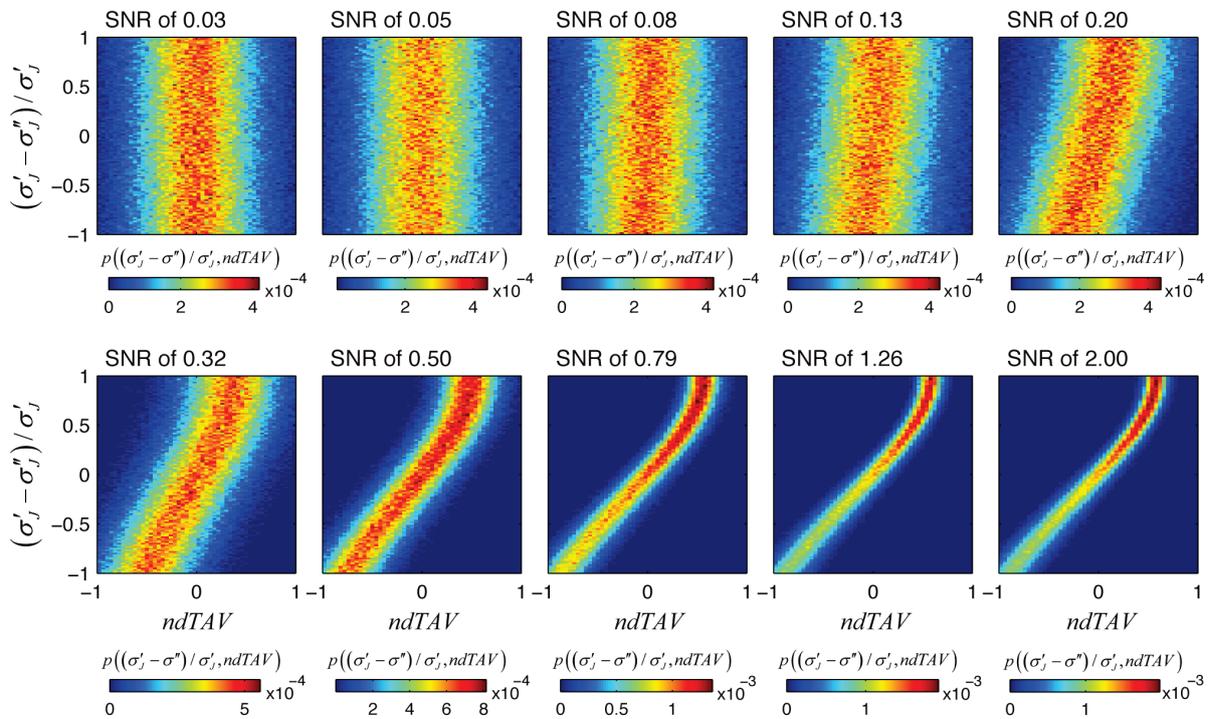

**Figure 6. Joint probability distribution of jitter reduction and $ndTAV$ for different SNR values for the mono-phasic neuronal response. An integration time window of $T_i$=300ms was used. For low SNR, $dTAV$ is uninformative as a measure of jitter reduction. As the SNR increases, $dTAV$ becomes more informative of the jitter reduction, i.e the ability to differentiate different levels of jitter reduction based on $dTAV$ improves substantially.**

## 3.2. Re-alignment of simulated data

We used the $dTAV$ and the MaxCorr realignment algorithms to realign neuronal responses in 100 simulated experiments for either mono-phasic or bi-phasic neuronal responses and for different levels of noise (Figure 7). To calculate $dTAV$ we used an integration window that captured the majority of the signal, starting at $T_S$ = 0s and ending at $T_E$ = 1s, both in respect to the stimulus times $t_E$. We intentionally did not want to use the results of the $dTAV$ reliability analysis (Figure 4d and 5d) to estimate the optimal integration window since this would bias our comparison with the MaxCorr algorithm. However, we did use the general finding of the $dTAV$ reliability analysis that using an integration window wider than the response diminished the $dTAV$ reliability less than using a window that is narrower than the response. In other words, we used a window that would certainly be wider than the optimal window, knowing that this does only weakly influence the reliability of $dTAV$ as a measure of jitter reduction. We used a filter window length of 250ms. For low SNRs (mono-phasic signal: SNR<0.2; bi-phasic signal: SNR<0.32) both algorithms increased the amount of jitter, rather than decreasing it. For intermediate SNRs (mono-phasic signal: 0.2<SNR<0.75; bi-phasic signal: 0.32<SNR<1.26; filter window 250ms) the $dTAV$ algorithm outperformed the MaxCorr algorithm ($p<10^{-9}$, Mann–Whitney–Wilcoxon signed test). For high SNRs (mono-phasic signal: 0.8<SNR; bi-phasic signal: 1.5<SNR), both algorithms removed almost all jitter from the recorded signal. In some of the high SNR cases, the MaxCorr algorithm achieved a higher jitter reduction, but the difference was very small (<0.08) and significant only in the case of the mono-phasic response for SNR=0.79 (P<0.05, Mann–Whitney–Wilcoxon signed test).

For both algorithms, the re-alignment worked best when a filter length of 250ms was used. For the $dTAV$ algorithm, the differences in the jitter reduction between using a filter window length of 250ms and using filter window lengths of 100ms and 500ms were very small. Only for a very long filter window (1000ms), the re-alignment performance decreased substantially. For the MaxCorr algorithm, the filter window length had a stronger influence on the amount of jitter reduction. When the filter window length is included as an additional parameter whose value was chosen by maximizing $dTAV$, the final jitter reduction was not significantly different from the jitter reduction that was obtained by the optimal filter length (250ms). Thus, the $dTAV$ algorithm can automatically select the proper filter length if filter length is included as one of the parameters.

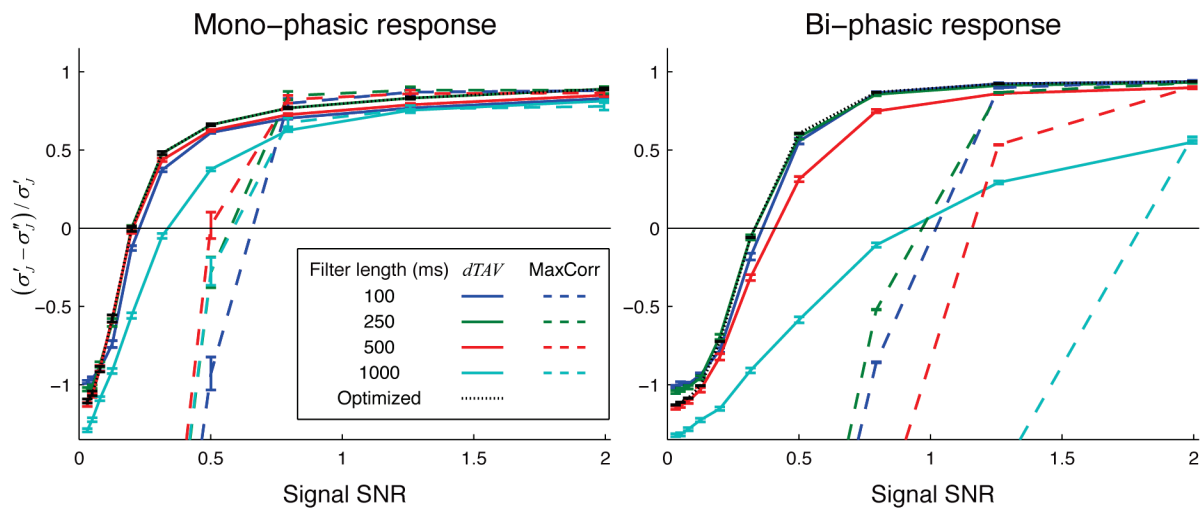

**Figure 7. Reduction of jitter standard deviation for the $dTAV$ algorithm (full lines) and the MaxCorr algorithm (dashed lines; [9]) for mono-phasic (left) and bi-phasic (right) neural responses. Reduction is shown for different low-pass filters and for the $dTAV$ algorithm when the filter length is optimized as one of the algorithm parameters (dotted black line). All results are averaged over 100 simulation repetitions; error bars depict the standard errors of the mean.**

# 4. Discussion

In this paper, we presented and evaluated a novel method for the realignment of neuronal responses. The algorithm uses the difference of time-averaged variance ($dTAV$) as a measure of jitter reduction, and the quadratic linear discriminant analysis to infer the temporal shifts of neuronal responses in individual trials. We showed that, by using the $dTAV$ algorithm, it is possible to realign simulated mono and bi-phasic single trial neuronal responses for noise levels several times larger than the neuronal response amplitudes.

We used mono and bi-phasic potentials composed out of one and two Gaussian functions as examples of neuronal responses. The success of the method in reducing the jitter for intermediate and low noise levels (SNR>0.32) and for both mono and bi-phasic neuronal responses, suggests that the algorithm can be successfully applied in a large number of cases. At the same time our results show differences between mono and bi-phasic response shapes, suggesting that the signal shape can affect the performance of the $dTAV$ algorithm.

Our simulations assumed that the recorded neuronal signal is a continuously modulated signal, which is valid for example for local field potentials, electro-corticographic signals (ECoG), electro-/magnetoencephalographic recordings (EEG/MEG) as well as for near-infrared spectroscopy (NIRS) and functional magnetic resonance imaging (fMRI) signals. Furthermore, our algorithm can also be applied to spike trains by estimating instantaneous neuronal firing from the spike times [15,16] and using the instantaneous neuronal firing rates as the neuronal signal in our algorithm. In addition, the algorithm can be applied to continuously modulated signals which were derived from the aforementioned neuronal signals, such as time-resolved spectral amplitudes (e.g. extracted using short-time Fourier transform) or crosscorelation measures. Recordings that depend on other variables than time, such as space or frequency, can also be realigned by exchanging the time variable by the corresponding variable (e.g. spatial coordinate or frequency).

We compared the performance of our $dTAV$ algorithm to the performance of the MaxCorr algorithm [9] which was also designed to reduce the temporal jitter in the neuronal responses. The $dTAV$ algorithm outperformed the MaxCorr algorithm substantially for intermediate SNRs and yielded similar performance for low and high SNRs.

The $dTAV$ algorithm works with a large number of parameters whose values have to be determined through testing different parameter sets. This can be achieved by defining admissible values for each parameter (e.g. within an interval) and scan all possible combinations. If the number of admissible values for each parameter is very high, this process becomes computationally demanding and, therefore, may be time consuming. In our simulations we obtained good realignments within a reasonable timeframe (less than an hour), by optimizing the parameters across a limited number of values. Indeed, our results show (e.g. Figures 4 and 5) that the fine optimization of parameters may not be necessary and the required computational time for parameter optimization may therefore not be problematic.

In general, realignment algorithms can be used to improve the analysis of neurophysiological experiments by improving the estimation of neuronal responses. The obvious case is the estimation of the neuronal response by calculating trial averages. As shown in our example (Figure 1), even if neuronal responses are constant across trials and the noise is uncorrelated to the signal, averaging the jittered single-trial responses can lead to a distorted estimation of the response and the incorrect estimation of the noise. Removing the jitter by the proposed algorithm can improve the estimation of the neuronal response and improves the estimation of the noise. A large number of neuroscience studies investigate neuronal responses related to sensory stimuli. When neuronal responses are well locked to the stimulus [17] realignment methods might be of limited use. On the other hand, neuronal responses may not be locked to the stimulus but, in addition to the stimulus, may also be affected by the internal neuronal state [18,19] and, therefore be temporally jittered relative to the stimulus onset. Our algorithm can be used to find out whether the responses are locked to such internal events and compute the approximate timing of these events. Furthermore, neuronal responses related to behaviour may also be jittered. For example, neuronal responses related to movement planning [20,21] may be jittered with respect to the times when the movements were initiated if the movements were either self-paced or triggered by another stimulus. Realignment algorithms could be used to align noisy individual trials in order to improve the accuracy of determining the underlying neuronal response (Figure 1). These more accurately determined neuronal responses would then also facilitate a comparison of responses between studies, allowing for identification of neuronal response parts shared between classes of responses. Computing the realignment times may give us also insights into the timing of internal events and into the variability of timings in internal cognitive processes.

Additionally, brain-machine interface systems that detect events based on neuronal recordings [22-29] may benefit from re-alignment algorithms. Such systems require a certain number of trials containing the neuronal responses to build the model used to detect the events from continuous neuronal recordings. If the jitter or the neuronal responses used to train the model is reduced, the detection may perform better.

In summary, we showed that the $dTAV$ realignment algorithm can reduce the jitter of simulated neuronal responses for response waveforms commonly observed in neurophysiological recordings and noise levels many times higher than the neuronal response itself. Hence, the application of the

$dTAV$ algorithm can improve analysis and interpretation of neuronal responses and improve the performance of asynchronous detection of events from neuronal recordings.

## Acknowledgements

This work was supported by the German Federal Ministry of Education and Research (BMBF) grant 01GQ0830 to BFNT Freiburg and Tübingen.